\documentclass[11pt]{article}
\usepackage{hyperref}
\usepackage{amsmath}
\usepackage{fullpage}
\usepackage[english]{babel}
\usepackage{titlesec}
\usepackage{graphicx}
\usepackage{subfig}
\usepackage{physics}
\usepackage{amsfonts}
\usepackage{subfig}
\usepackage{float}
\usepackage{epsfig}
\usepackage{setspace}
\usepackage{authblk}
\DeclareMathOperator{\e}{e}

\begin{document}

\title{Quantum correlations under the effect of a thermal environment in a triangular optomechanical cavity.}

\author[1]{Oumayma El Bir \thanks{oumayma.elbir@um5s.net.ma}}
\author[1,2]{Morad El Baz \thanks{morad.elbaz@um5.ac.ma} }
\affil[1]{ESMaR, Faculty of Sciences, Mohammed V University, Rabat, Morocco}
\affil[2]{The Abdus Salam International Center for Theoretical Physics, ICTP-Trieste, Italy}
\date{\today}
\maketitle

\begin{abstract}
We quantify the stationary correlations between the optical mode and the relative mechanical mode of a ring cavity composed of a fixed mirror and two movable ones in a triangular design. The bipartite covariance matrix, is used to evaluate the logarithmic negativity as a measure of entanglement, the Gaussian quantum discord as a measure of total quantum correlations and the mutual information as a measure of the overall correlations. The behaviour of these quantities with respect to the environment's temperature as well as other parameters such as the laser pumping power and mass of the movable mirrors is discussed.

\end{abstract}

\section{Introduction}
In recent years, quantum information has attracted a lot of attention because of some of its surprising, unexpected, and remarkable properties. One of the most interesting characteristics at the foundations of this field is "entanglement" \cite{20}, which is, specifically, a  quantum property. So far, the search on the production and manipulation of entangled states has been achieved with success in microscopic systems such as, atoms, photons, ions,... \cite{21}, \cite{22}, \cite{23}. Whereas, the validity of entanglement in the macroscopic domain is not yet completely clear, this question made it an interesting subject for physicists.  In this context, optomechanical systems manifest as a new platform to investigate the entanglement between a macroscopic mechanical object and an optical field.

The optomechanical systems \cite{1} are typically composed by an optical cavity with a moving mechanical object which can interact with the electromagnetic field confined in the cavity via the radiation pressure force \cite{24}. Indeed this latter is due to the impulse transfer of the photons during their reflection on the surface of the mechanical object. This force induces small displacements of the motion of the mechanical object which change the length of the cavity, and modify the state of the cavity field. This can be well observed when the cavity is driven by a strong laser source \cite{25}. Such effect is called the optomechanical coupling which designs the light-matter interaction.

In this paper, we study the quantum correlations that are present in a ring cavity with two movable mirrors, the motion of which, is due to the effect of the radiation pressure force. This allows to model the mirrors' motion as a mechanical harmonic oscillators and thus can be considered as a continuous variable system. We give the degree of entanglement by using the logarithmic negativity, we describe also the quantum mutual information which gives the total amount of correlations contained in a quantum state which is equal to the sum of classical correlations and the quantum discord \cite{26, 27}. In fact, quantum discord has drawn much attention recently as another measure of quantum correlations that differs from entanglement, as the latter can be seen as contained in the former. Indeed quantum discord was first presented by \cite{28, 29}, where it has been proved that some separable states do contain some amount of quantum correlations and have a nonzero value of the quantum discord. This notion was generalized to the field of continuous variables (CV) systems in references \cite{30, 31}.

The structure of this work is organized as follows. In Sec II we describe the model and present the Hamiltonian, then we solve the nonlinear quantum  Langevin equations for the optical modes and the relative quadratures of the two mechanical modes which leads to obtain the bipartite covariance matrix. In Sec. III we study the stationary entanglement by using the logarithmic negativity, and discuss the effect of some parameters on it. Beyond entanglement we study the Gaussian quantum discord in Sec  IV to measure the quantum correlations especially in the case where the state is separable. We simulate also the mutual information to quantify the total correlations present in the system considered. Finally we conclude by some remarks to close the paper.
\section{System and model}
\subsection{System}
\begin{figure}[ht]
\centering
\includegraphics[scale=2]{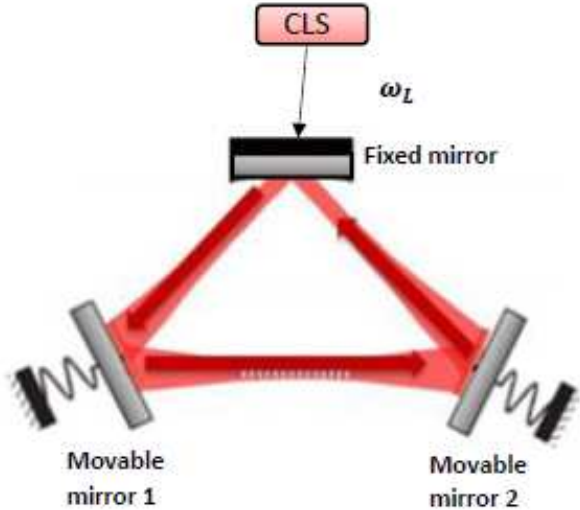}
\caption{The scheme of the optomechanical system under study including a ring cavity pumped by a coherent laser source (CLS) with frequency $\omega_{L}$.} 
\label{ring}
\end{figure}

We consider the ring cavity schematized in Figure \ref{ring}, with arm length $L$ and frequency $\omega_c$. This type of cavity was introduced in many different ways \cite{0, v} and further studied in \cite{2} The cavity  has three mirrors in a triangular configuration, such that one of the mirrors is fixed and partially transmitting, whereas the other two are movables and perfectly reflecting. These moving mirrors are considered as quantum harmonic oscillators with frequency respectively $\omega_{m1}$, $\omega_{m2}$ and effective masses $m_{1}$, $m_{2}$. Their loss of mechanical energy is described by the damping rates $\gamma_{m1}$, $\gamma_{m2}$, respectively. The cavity is pumped by a coherent laser field with frequency $\omega_L$  and the mean amplitude of the input laser is given by  $E = \sqrt{\frac{2 \kappa P}{\hbar \omega_L}}$ where $P$ is the input power pump,  $\hbar$ is Planck constant and $\kappa$ is  the total optical decay rate of the cavity: $\kappa= \kappa_{ex} + \kappa_0$ \cite{1}. Here $\kappa_{ex}$ denotes the losses at the input of the cavity and $\kappa_0$ is due to internal losses.

An important parameter is the optical finesse $F$, which gives the average number of round-trips before a photon leaves the cavity: 
\begin{equation}
\begin{aligned}
F = \frac{\pi c}{\kappa L},
\end{aligned}
\end{equation}
c denotes the speed of light.

The reflection of each photon on the movable mirror gives rise to transfer of momentum equal to $2 \hbar k$, where $k$ is the wave vector of the field. This transfer produces a radiation pressure force $F_{rad}(t)$ that is exerted on the mirror. This force is equal to the momentum exchanged per photon, multiplied by the number of photons reflected per second:
\begin{equation}
\begin{aligned}
\hat{F}_{rad} \simeq 2 \hbar k \hat{I}.
\end{aligned}
\end{equation}
This force induces a displacement on the movable mirrors, that makes it possible to have a correlation between the position of the mirrors and the intensity of the radiation $\hat{I}$.

\subsection{The Hamiltonian}

The Hamiltonian of the system is given by \cite{2}
\begin{equation}
\begin{aligned}
H = \hbar \Delta_{0} a^{+}a + \frac{\hbar \omega_{m1}}{2}(q_{1}^{2} + p_{1}^{2}) + \frac{\hbar \omega_{m2}}{2}(q_{2}^{2} + p_{2}^{2}) + \hbar g a^{+}a \cos^{2}(\theta /2)(q_{1} - q_{2}) + i \hbar E (a^{+} - a), 
\end{aligned}
\label{eq:2}
\end{equation}
where $a$ and $a^{+}$ are respectively, the annihilation and creation operators of the optical mode, with $[a, a^{+}] = 1$ and $\Delta_{0} = \omega_{c} - \omega_{L}$ denotes the detuning between the cavity and the laser radiation.
The first term in (\ref{eq:2}) describes the energy of the cavity mode while the second and third terms describe the energy of the mechanical resonator. We defined $q_{i}$ and $p_{i}(i = 1,2) $ as being, respectively, the dimensionless position and momentum operators of the oscillator  with $[q_{j} , p_{k}] = i \delta_{jk} \ (j, k = 1, 2)$. The fourth term denotes the  interaction between the cavity field and the movable mirrors which is due to the radiation pressure, g is the optomechanical coefficient given by $g = \frac{\omega_{c}}{L} \sqrt{\frac{\hbar}{m \omega_{m}}}$ in units of $s^{-1}$.  The last term describes the laser driving input.  

\subsection{The quantum Langevin equations}
The analysis of the dynamics of the two coupled relative mechanical modes and the cavity field are determined by the Heisenberg equations of motion, which are derived from (\ref{eq:2}), and by adding the fluctuation-dissipation terms, we obtain the following set of Heisenberg-Langevin equation 
\begin{equation}
\begin{aligned}
\dot{q_{1}} &= \omega_{m_{1}} \ p_{1}, \\
\dot{q_{2}} &= \omega_{m_{2}} \ p_{2}, \\
\dot{p_{1}} &= -g a^{+}a \cos^{2}(\theta /2) - \omega_{m_{1}} q_{1} - \gamma_{m1} \ p_{1} + f_{1}, \\
\dot{p_{2}} &= \ g a^{+}a \cos^{2}(\theta /2) - \omega_{m_{2}} q_{2} - \gamma_{m2} \ p_{2} + f_{2},\\
\dot{a} &= -i[\Delta_{0} + g \cos^{2}(\theta /2)(q_{1} - q_{2})]a + E - \kappa a + \sqrt{2 \kappa} a_{in},\\
\dot{a}^+ &= i[\Delta_{0} + g \cos^{2}(\theta /2)(q_{1} - q_{2})]a^+ + E - \kappa a^+ + \sqrt{2 \kappa} a_{in}^+,
\end{aligned}
\end{equation}
We have introduced the input vacuum noise $a_{in}$ at temperature $T_{i}$ with nonzero time domain correlation functions \cite{4}:
\begin{equation}
\begin{aligned}
<a_{in}(t) \ a_{in}^+(t^{'})> = \delta(t-t^{'}).
\end{aligned}
\end{equation}
In addition to that, we have the Brownian noise expressed by the operator $f_{i}(i=1, 2)$, which describes the coupling of the $i^{th}$ movable mirror to it's own environment, with zero mean values and characterized by the following correlation functions \cite{05}:

\begin{equation}
\begin{aligned}
<f_{i}(t) f_{j}(t^{'})> = \frac{\delta_{jk}}{2 \pi} \frac{\gamma_{m}}{\omega_{m}} \int \omega \e^{- i \omega (t - t^{'})} [1 + \coth(\frac{\hbar \omega}{2 k_{B} T})] d\omega,
\end{aligned}
\end{equation}
The quality factor obeying $Q_{i} = \frac{\omega_{m_{i}}}{\gamma_{m_{i}}} \gg 1$ means that one can assume that the mechanical baths are Markovian. Hence the $f_{i}$ non-zero correlation functions \cite{5} are given by   
\begin{equation}
\begin{aligned}
<f_{i}(t) f_{i}(t^{'}) + f_{i}(t^{'}) f_{i}(t)>/2 \ \simeq \gamma_{m} (2 n_{th_{i}} + 1)\delta(t - t^{'}),
\end{aligned}
\end{equation}
Where $n_{th_{i}} = [\e^{(\frac{\hbar \omega_{m_{i}} }{k_{B} T_{i}})} - 1]^{-1}$ is the $i^{th}$ thermal photon number and $k_{B}$ is the Boltzmann constant, $i,j = 1,2$.\\
Without loss of generality, we choose $\omega_{m_{1}} = \omega_{m_{2}} = \omega_{m}$, $\gamma_{m_{1}} = \gamma_{m_{2}} = \gamma_{m}$, $m_{1} = m_{2} = m$  and $T_{1} = T_{2} = T ( n_{th_{1}} =  n_{th_{2}} =  n_{th})$.

Introducing the relative distance: $q_{-} = q_{1} - q_{2}$ and the relative momentum: $p_{-} = p_{1} - p_{2}$ of the movable mirrors, we obtain the reduced equations
\begin{equation}
\begin{aligned}
\dot{q} &= \omega_{m} \ p, \\
\dot{p} &= -2  g   n_{cav} \cos^{2}(\theta /2) - \omega_{m} q - \gamma_{m} \ p + f_{1} - f_{2}, \\
\dot{a} &= -i[\Delta_{0} + g \cos^{2}(\theta /2)q]a + E - \kappa a + \sqrt{2 \kappa} a_{in},\\
\dot{a}^+ &= i[\Delta_{0} + g \cos^{2}(\theta /2)q]a^+ + E - \kappa a^+ + \sqrt{2 \kappa} a_{in},
\label{eq:8}
\end{aligned}
\end{equation}
 where $n_{cav}$ is the averge number of the photons inside the cavity.
 
\subsection{Linearization of the quantum Langevin equations}

To study the quantum correlations that are present in the system, we have to calculate the fluctuations. These can be obtained by linearizing the quantum Langevin equations (\ref{eq:8}). This is achieved by expanding each Heisenberg operator as a sum of its c-number classical steady-state value plus an additional small fluctuation operator with zero-mean value \cite{6}:
\begin{equation}
\begin{aligned}
a = \alpha_{s} + \delta a, \  q = q_{s} + \delta q, \  p = p_{s} + \delta p .
\label{xxx}
\end{aligned}
\end{equation}
The steady-state values are given by the following nonlinear algebraic equations: 
\begin{eqnarray}
p_{s} &=& 0, \nonumber \\
(i \Delta + \kappa )\alpha_{s} - E &=& 0, \nonumber \\
q_{s} + \frac{2 g \cos^{2} (\theta /2) |\alpha_{s}|^2 }{\omega_{m}} &=& 0,
\end{eqnarray}
where $\Delta = \Delta_{0} + g \ q_{s} \cos^{2} (\theta /2)$ is the effective cavity detuning.

Inserting equation (\ref{xxx}) into equation (\ref{eq:8}), and introducing the cavity field quadratures $\delta x = \delta a + \delta a^+$
and $\delta y = i(\delta a^+ - \delta a)$, and the input noise quadratures $\delta x_{in} = a_{in} + a_{in}^+$ and $\delta y_{in} = i ( a_{in}^+ - a_{in}) $, allows to obtain the linearized quantum Langevin equations:
\begin{equation}
\begin{aligned}
\delta\dot{q} &= \omega_{m} \ \delta p, \\
\delta \dot{p} &= -\omega_{m} \delta q - \gamma_{m} \ \delta p  - g \cos^{2}(\theta /2)(\alpha_{s}^* + \alpha_{s} ) \delta x + i g \cos^{2}(\theta /2)( \alpha_{s} - \alpha_{s}^* )\delta y + f_{1} - f_{2}, \\
\delta\dot{x} &= -i g \cos^{2}(\theta /2)( \alpha_{s} - \alpha_{s}^* )\delta q - \kappa \delta x + \Delta \delta y + \sqrt{2 \kappa} \delta x_{in}, \\
\delta\dot{y} &= - g \cos^{2}(\theta /2)( \alpha_{s}^* + \alpha_{s} )\delta q - \kappa \delta x - \Delta \delta y + \sqrt{2 \kappa} \delta y_{in}.
\label{11}
\end{aligned}
\end{equation}
This system of equations can be rewritten in the matrix form
\begin{equation}
\dot{u}(t) = A u(t) + \eta(t),
\label{xx}
\end{equation}
where $ u^{T}(\infty) = (\delta q(\infty), \delta p(\infty), \delta x(\infty), \delta y(\infty))$ is the column vector of the fluctuations, $ \eta^T(t) = (0, f_{1} - f_{2}, \sqrt{2 \kappa} \delta x_{in}, \sqrt{2 \kappa} \delta y_{in}) $ is the column vector of noise operators and the drift matrix A reads 
\begin{equation}
A = \begin{pmatrix} 0 & \omega_{m} & 	0 & 0 \\ - \omega_{m} &  - \gamma_{m} & -2 g \cos^{2}(\theta /2)Re[\alpha_{s}] & 2 i g \cos^{2}(\theta /2)Im[\alpha_{s}]\\ -2 i g \cos^{2}(\theta /2)Im[\alpha_{s}] & 0 & - \kappa & \Delta \\
- 2 g \cos^{2}(\theta /2)Re[\alpha_{s}] & 0 & - \Delta & - \kappa \end{pmatrix}
\end{equation}
where $Im[\alpha_{s}]$ ($resp$ $Re[\alpha_{s}]$) denotes the Imaginary ($resp$ real) part of $\alpha_{s}$. The solution of equation \ref{xx} is $u(t) = Y(t)u(0) + \int_{0}^{t} dx Y(x) \eta(t - x)  $, where $Y(t) = \exp{At}$. 

\subsection{Covariance Matrix}

When the stability conditions are verified according to the Routh-Hurwitz criteria \cite{7} \cite{8} \cite{9}, the steady-state Covariance Matrix satisfies the Lyapunov equation:
\begin{equation}
\begin{aligned}
A V + V A^T = -D,
\label{18}
\end{aligned}
\end{equation}
where D is a diagonal matrix that represents the noise correlations; it is given by $D = Diag[0, \gamma_{m}(2 n_{th} + 1), \kappa, \kappa]$.

In the following we express the Covariance Matrix V  in a 2 x 2 block form:
\begin{equation}
V = \begin{pmatrix} V_{m} & V_{c} \\ V_{c}^T & V_{a} \end{pmatrix},
\end{equation}
where $V_{a}$, $V_{m}$ and $V_{c}$ are respectively, the Covariance Matrix of the optical cavity mode, the relative mechanical mode and the non-local correlations between them.

For later use, We define the following set of 4 symplectic invariants derived from the Covariance Matrix V
\begin{equation}
\begin{aligned}
I_{1} &= Det \ V_{m} \ \ \ \ , \ \ \ \ I_{2} &= Det \ V_{a} \\
I_{3} &= Det \ V_{c} \ \ \ \ , \ \ \ \ I_{4} &= Det  \ V.
\end{aligned}
\end{equation}
For the bipartite system the symplectic eigenvalues of the partial transpose of the Covariance Matrix V \cite{10} \cite{11} are given by
\begin{equation}
\begin{aligned}
\tilde{\nu}_{\pm}^2 = \sqrt{\frac{\tilde{\Gamma} \pm  \sqrt{\tilde{\Gamma}^2 - 4 I_{4}}}{2}},
\label{33}
\end{aligned}
\end{equation}
where the symbol $\tilde{\Gamma}$ is the symplectic invariant given by $\tilde{\Gamma} = I_{1} + I_{2} - 2 I_{3}$.

Knowledge of the Covariance Matrix in the stationary state from (\ref{18}) and use of the quantities \ref{33}, allows to evaluate the quantum correlations between the optical mode and the relative mechanical mode. This is studied in the next section.
\section{Entanglement analysis}
In this part we want to quantify the bipartite entanglement between the mechanical mode and the optical mode, for that we use the logarithmic negativity $E_{N}$. This quantity is defined for Gaussian continuous variable (CV) systems as \cite{12, 13}
\begin{equation}
\begin{aligned}
E_{N} = \max [0, -Ln \ 2\tilde{\nu}_{-}].
\end{aligned}
\end{equation}
The two modes are entangled when $E_{N} > 0$, which is equivalent to $\tilde{\nu}_{-} <  \frac{1}{2}$. If not the states are separable, which is consistent with Simon's necessary and sufficient criterion for bipartite entanglement \cite{15}.

In order to study the behaviour of the quantum correlations, as captured by the logarithmic negativity, we use parameters taken from the most relevant experiments \cite{16}, where the length of the cavity $L = 25 mm$, the laser wave length is $\lambda = 1064 nm$, $\omega_{m} = 2 \pi .947.10^3 H_{z}$ and the mechanical quality factor $ Q = \frac{\omega_{m}}{\gamma_{m}} = 6700 $. \\
\begin{figure}[h]
\centering
\includegraphics[scale=1.2]{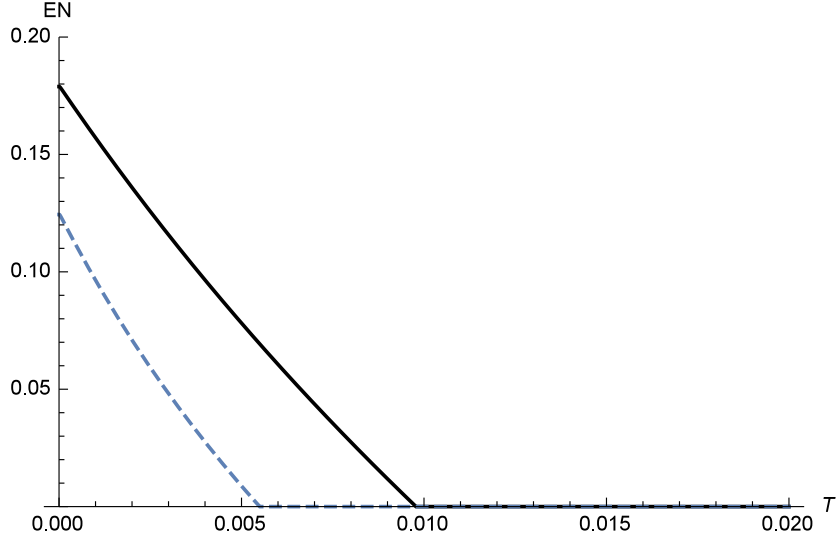}
\caption{Plot of the logarithmic negativity $E_{N}$ versus the thermal bath temperature $T(K)$, for two values of the mass. $m = 50 ng$ for the black full line and $m = 100 ng$ for the blue dashed line. The other parameters are chosen as follows: $\Delta = \omega_{m}$, $P = 3.8 \ mW$,  and $\kappa = 2 \pi . 215 . 10^3 \ H_{z}$.}
\label{zz}
\end{figure}
Figure \ref{zz} shows the logarithmic Negativity $E_{N}$ versus the thermal bath's temperature T for different values of m. We notice that, as expected, the quantum correlations do not resist against the environment effects as $E_{N}$ decreases with temperature, and that for $m = 50 \; ng$, it survives until $T > 9 \; mK$ while for $m = 100 \; ng$ it vanishes at $T = 6 \; mK$. On the other hand, for a fixed temperature the logarithmic negativity decreases with mass {\it{e.g.}}  when $T = 3 \; mK$, $E_{N} = 0.11 \; (\hbox{respectively } 0.04)$ for $m = 50 \; ng \; (\hbox{respectively } 100 \; ng)  $, which shows the importance of the parameter mass as well. As a matter of fact, the smaller it is, the more robust the system becomes against the thermal bath's temperature. This in turn is due to the fact that the optomechanical coupling, due to the radiation pressure, gets stronger as the displacement of the mirrors increases.

\begin{figure}%
    \centering
    \subfloat[]{{\includegraphics[width=8cm]{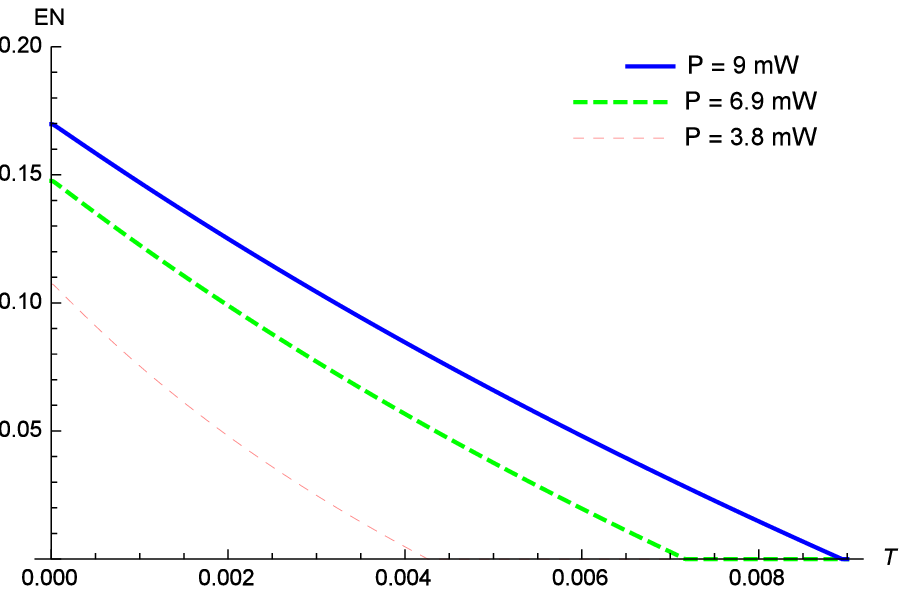}\label{111} }}%
    \,
    \subfloat[]{{\includegraphics[width=8cm]{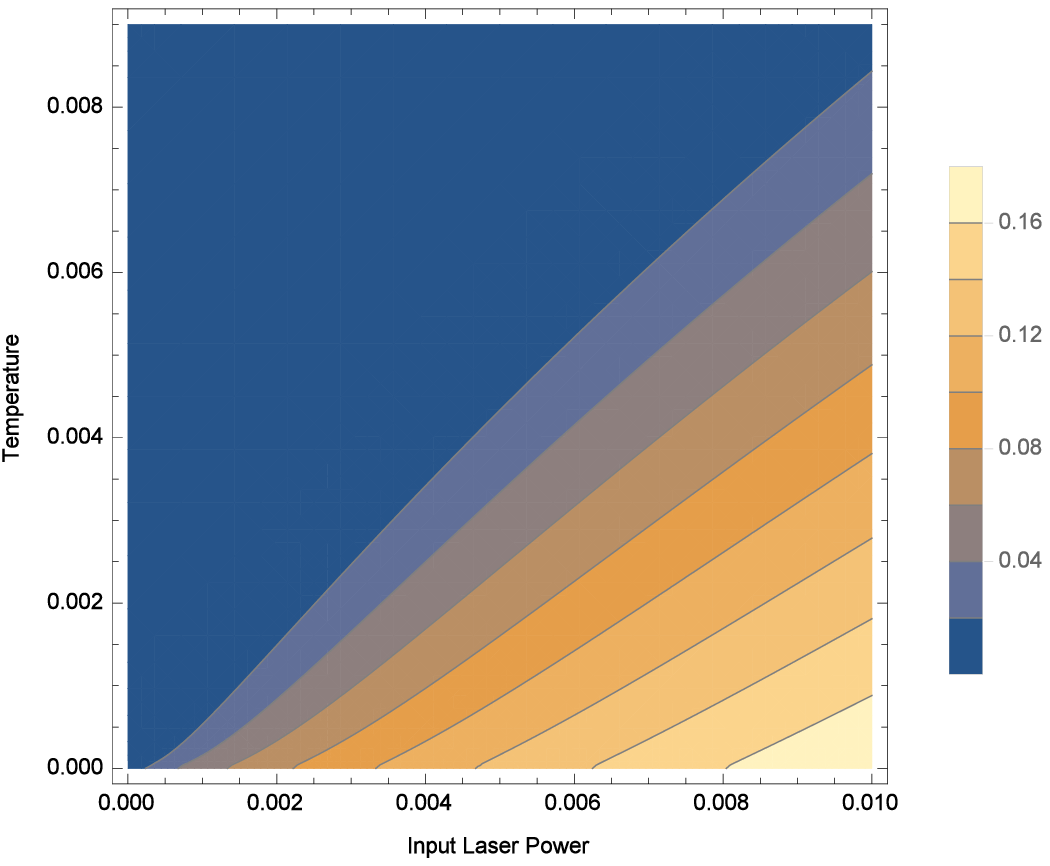} }}%
    \caption{(a) shows $E_{N}$ versus $T$ for different values of the input laser power P. (b) shows the CountourPlot of $E_{N}$ vs the temperature T and the power P. In both plots we use  $m = 145 ng $, $\Delta = 0.965 \omega_{m}$ and $\kappa = 2\pi.215.10^3$. \label{essay}}%
\end{figure}

Next, to examine the dependence of entanglement on some other parameters in an optomechanical system, we show in Figure (\ref{111}) its behaviour as a function of the laser driving power P as well. We notice, for instance, that for fixed temperature, the driving laser power has a favorable effect on the quantum correlations as captured by the logarithmic negativity. For example at $T = 2 \;  (mK)$, we obtain $E_{N} = 0.048$ for  $P = 3.8 \; mW $, $E_{N} = 0.098$ for $P = 6.9 \; mW $  and $E_{N} = 0.125$ for $P = 9 \; mW $.  We also see from the countour plot, that the temperature, at which the entanglement vanishes, increases with the power, {\it{e.g.}} for $P = 4 \; mW$, $E_{N}$ vanishes at $T \sim 4.5 \; mK$ however for $P = 10 \; mW$, $E_{N}$ persists up to $ 9 \; mK$.



\section{Gaussian quantum discord and the quantum mutual information}
Basing solely on the entanglement, we can conclude that the states of the two modes are entangled if $E_{N} > 0$ otherwise the states are separable or classically correlated. Nevertheless, it was proved \cite{se, st} that even separable states may contain quantum correlations, thus the notion of quantum discord was introduced as capturing the overall quantum correlations, even those beyond entanglement. Indeed in this section, we use the Gaussian quantum discord, \cite{30, 19} which is considered as a measure of nonclassical correlations in Gaussian quantum systems that can be different from zero even for separable states {\it{i.e.}} $E_{N} = 0$. By definition the quantum discord is the difference between the mutual information $I_{M}$ which is a measure of the total correlation in the system and the classical correlation $C$ :
\begin{equation}
\begin{aligned}
D = I_{M} - C.
\end{aligned}
\end{equation}

\begin{figure}[h]
\centering
\includegraphics[scale=1]{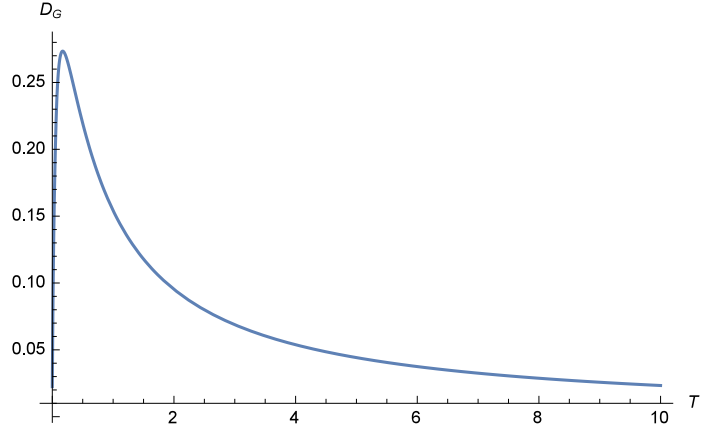}
\caption{Plot of the Gaussian quantum Discord vs the temperature T. We chose $P = 3.8 \; mW$, $\Delta = \omega_{m}$, the other parameters are similar to those in Figure \ref{essay}.}
\label{gf}
\end{figure}

The Gaussian quantum discord for Gaussian bipartite systems can be expressed as \cite{30, 19}
\begin{equation}
\begin{aligned}
D_{G} = g(\sqrt{I_{2}}) - g(\nu_{+}) - g(\nu_{-}) + g(\sqrt{W}),
\end{aligned}
\end{equation}
where, g is a function defined by
\begin{equation}
\begin{aligned}
g(x) = (x + \frac{1}{2})Ln(x + \frac{1}{2}) - (x - \frac{1}{2})Ln(x - \frac{1}{2}),  
\end{aligned}
\end{equation}
and 
\begin{equation}
    W = \left\{
      \begin{aligned}
        & \frac{2 |I_{3}| + \sqrt{4 I_{3}^2 + (4 I_{2} - 1)(4 I_{4} - I_{1})}}{(4 I_{2} - 1)}^2   \ \ \ \ \ \ \ \ \ \ \ \ \ \ \ \ \ \ \ \ \ \ \hbox{if } \frac{4(I_{1}I_{2} - I_{4})^2}{(I_{1} + 4 I_{4})(1 + 4I_{2}) I_{3}^2} \leq 1 \\
        &\frac{I_{1}I_{2} + I_{4} - I_{3}^2 - \sqrt{(I_{1} I_{2} + I_{4} - I_{3}^2)^2 - 4 I_{1} I_{2} I_{4}} }{2 I_{2}} \ \ \ \ \ \ \ \ \ \ \ \ \ \ \ \ \ \  \hbox{otherwise}
      \end{aligned}
    \right.
\end{equation} \\
and $\nu_{\pm}$ are the symplectic eigenvalues of $V$:
\begin{equation}
\begin{aligned}
\nu_{\pm} = \frac{1}{2} \sqrt{\Gamma^\pm  \sqrt{\Gamma^2 - 4 I_{4}}},
\end{aligned}
\end{equation}
with $\Gamma = I_{1} + I_{2} + 2I_{3}$.

Figure \ref{gf} showing the Gaussian quantum discord versus the temperature confirms that the discord decreases as $T$ increases. Furthermore, we note that while $E_{N}$ vanishes at $T \sim 5$ mK the Gaussian quantum discord survives beyond this temperature. This explains the existence of the quantum correlations in the system even for separable states and also shows the robustness of the Gaussian quantum discord against the influence of the environment.

While the quantum correlations are captured by the Gaussian quantum discord, the total correlations can be evaluated by using the expression of the quantum mutual information \cite{19}: 
\begin{equation}
\begin{aligned}
I_{M} = g(\sqrt{I_{1}}) + g(\sqrt{I_{2}}) - g(\nu_{+}) - g(\nu_{-}).
\end{aligned}
\end{equation}

\begin{figure}%
    \centering
    \subfloat[The mutual information as a function of the normalized effective detuning $\Delta$ / $\omega_{m}$.]{{\includegraphics[width=8cm]{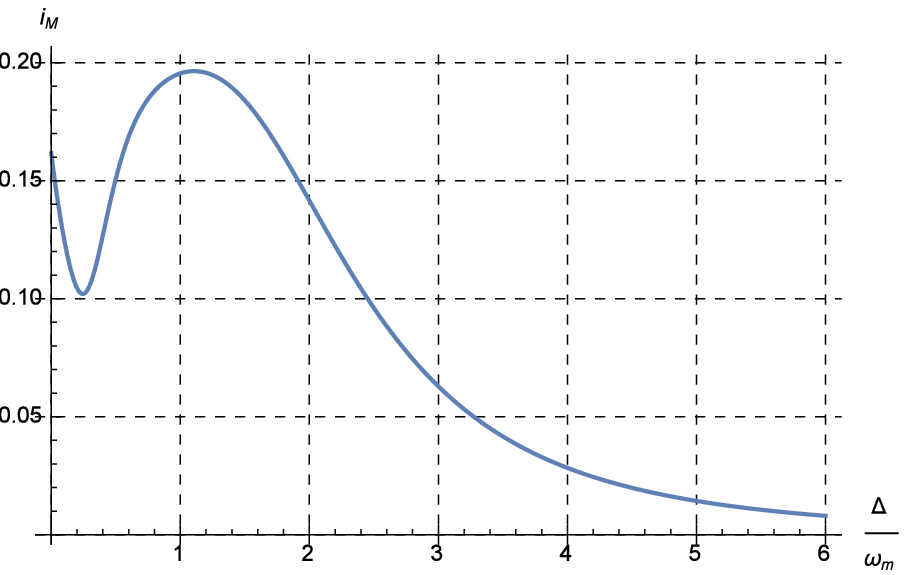} }}%
    \,
    \subfloat[The Gaussian quantum Discord as a function of the normalized effective detuning $\Delta$ / $\omega_{m}$.]{{\includegraphics[width=8cm]{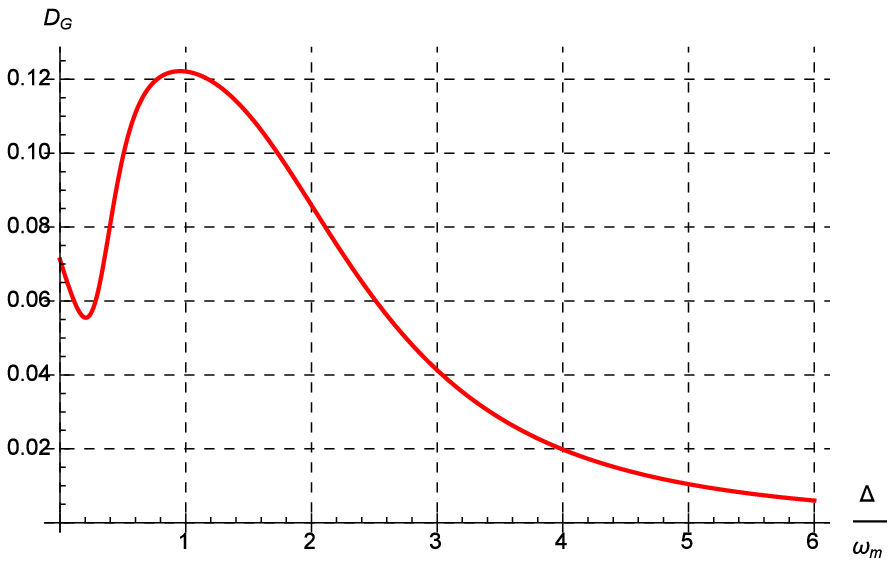} }}%
    \caption{The mutual information and The Gaussian quantum Discord. In both plots the temperature of the environment is taken $T = 6 \; mK$ with the other parameters similar to those in Figure \ref{gf}. \label{sss}}%
\end{figure}


Figures \ref{sss}, show respectively, the behaviour of the mutual information $I_{M}$ and that of the quantum Gaussian discord $D_{G}$ versus $\Delta$ / $\omega_{m}$. The two quantities behave similarly and reach their maximum in the vicinity of $\Delta \sim \omega_{m}$ then they decrease with increasing of $\Delta$ / $\omega_{m}$. Furthermore, for fixed values of $\Delta $ / $\omega_{m}$ we have $I_{M}$ always greater than $D_{G}$ which is logical since $I_{M}$ measures both the classical and the quantum correlations present in the system.
\section{conclusion}
In this work, we have studied the quantum correlations in a steady state of an optomechanical system. In our case, this latter is a ring cavity, with a fixed mirror and two movable ones, pumped by a laser source. The linearized quantum Langevin equations were used in order to obtain the bipartite Covariance Matrix at the stationary state fully describing the Gaussian state of the two subsystems, the optical modes and the relative mechanical mode. 

In addition to that, we have studied the effect of the thermal bath temperature on the quantum entanglement which is measured by the logarithmic negativity. It was found that the entanglement decreases with increasing temperature and mass of the two mechanical harmonic oscillators.

Then, the total amount of correlations present in the system under consideration were quantified by the quantum mutual information, a measure englobing both the classical correlations and the quantum discord. It was found that both the quantum mutual information and the quantum discord decrease with increasing $\Delta / \omega_{m}$ and they reach their maximum when $\Delta$ is around $\omega_{m}$. As expected when the entanglement vanishes (which is characteristic of a separable state) the system still has a non-zero value of the Gaussian quantum discord. This shows that the Gaussian quantum discord is a measure of all quantum correlations, including entanglement and confirms the robustness of this measure and the feebleness of the logarithmic negativity against the fluctuations of the bath environment.

\end{document}